# Local stress and elastic properties of lipid membranes obtained from elastic energy variation

Maksim A. Kalutsky, Timur R. Galimzyanov, Konstantin V. Pinigin*

A.N. Frumkin Institute of Physical Chemistry and Electrochemistry, Russian Academy of Sciences, 31/4 Leninskiy prospekt, Moscow 119071, Russia

*Email: piniginkv@gmail.com

ABSTRACT: A theory and computational method are provided for the calculation of lipid membranes elastic parameters, which overcomes the difficulties of the existing approaches and can be applied not only to single-component but also to multi-component membranes. It is shown that the major elastic parameters can be determined as the derivatives of the stress-profile moments with respect to stretching. The more general assumption of the global incompressibility, instead of the local one, is employed, which allows the measurement of the local Poisson's ratio from the response of the stress profile to the isotropic ambient pressure. In the case of the local incompressibility and quadratic energy law, a direct relation between the bending modulus and Gaussian curvature modulus is established.

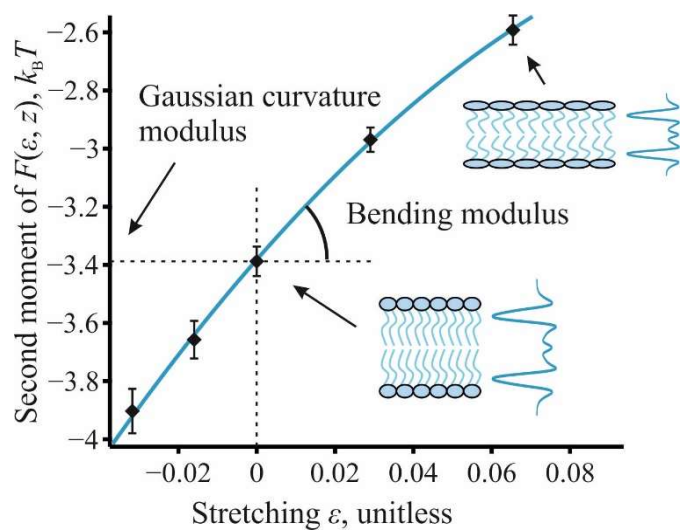

# Introduction.

The biological role of lipid membranes is crucial in living cells. Although their primary function is providing a shell for cells and their compartments, lipid membranes are also involved in such important processes as signaling, trafficking, metabolism, fusion and fission, exo- and endocytosis [1]. The principal features of these processes, such as characteristic times and energy barriers, are controlled by membrane reshaping, which makes the mechanical properties of lipid membranes a substantial factor in the life cycle of individual living cells and organisms.

To account for the membrane mechanics the theory of elasticity is often employed [2–8]. In this framework, the energetic response to the distorting forces is quantified by the elastic moduli of corresponding deformations. The values of these moduli are one of the key elements for the description of the membrane-involved processes. Usually, four main deformations are used to characterize membrane reshaping: bending (splay), tilt, stretching-compression, and saddle (Gaussian) splay. Large-scale membrane deformations can be described only by the bending mode. This description leads to the classical Helfrich Hamiltonian [3]. At a smaller scale, comparable to the membrane thickness, stretching and tilt modes should also be taken into account. Another important parameter is spontaneous curvature which describes the preferred bending state of the membrane or individual monolayers caused by an internal pre-stress in the monolayers or membrane asymmetry.

The classical approach treats elastic moduli as phenomenological ones, determined in experiments or molecular dynamics (MD) simulations. Each way comprises a wide variety of methods for the determination of membrane elastic parameters. MD approach provides full and precise control of the system's conditions and composition unattainable in the experimental setup, making possible a thorough investigation of the elastic moduli and analysis of various effects at small scales and extremely high curvatures. Despite the great advances in MD methods, there are still enough complications in their application. A large and important part of these methods relies on monitoring fluctuations [9–19] of either a membrane surface or orientations of individual lipids, called directors. A recent comprehensive study shows that the ambiguities in the definitions of directors lead to substantial systematic errors that may be as large as 32% for the bending modulus of pure lipid systems, and one should also be careful with the definition of the membrane surface [19]. The situation with lipid mixtures is conceivably much worse. Furthermore, the consideration of lipid mixtures demands assumptions, the validity of which is rather vague, such as an analog of the Reuss averaging for the bending modulus employed in Ref. [11]. Another issue with the fluctuation methods is that it is impossible to determine monolayer spontaneous curvature, as it does not manifest itself in the fluctuations due to periodic boundary conditions. The other methods are based on the application of external forces on the membrane and the measurement of the deformation response. For example, the stretching modulus can be determined by monitoring the area per lipid at different tensions [20]. The bending modulus though requires a more complicated procedure. One of the proposed methods [21,22] is based on membrane buckling during compression. Although this approach has the advantage of not assuming any microscopic underpinning of the Helfrich Hamiltonian, it requires a large lipid system to be simulated and is therefore out of reach for all-atom (AA) simulations. Besides, this method does not apply to lipid mixtures due to inevitable composition-curvature coupling and is not applicable for the determination of spontaneous curvature. An alternative approach for the determination of the bending modulus involves simulating tethers [23]. This method has technical limitations related to the requirement to balance pressure and area differences. Also, the latter approach does not provide spontaneous curvature values. Another direct force approach relies on the parabolic-like bending of the membrane system [24]. Like the buckling protocol, the latter does not apply to lipid mixtures and does not provide the spontaneous curvature values. Also, this method should be used with great caution since the considered system does not possess constant curvature. Imprecise assumptions may lead to false conclusions like the strong dependence of the bending rigidity on the membrane curvature. Yet in Ref. [25] it has been shown that a thorough description of the shape, which applies the Helfrich energy functional to each monolayer separately, restores the constancy of the bending rigidity as a function of curvature.

In this paper, we consider the local stress profile approach that on the one hand was previously largely unconsidered, but on the other hand, is widely used in other methods in one of its aspects. At the same time, we show that it has the potential to overcome the difficulties of the existing methods for the measurement of membrane elastic characteristics. This approach is already broadly employed for the determination of monolayer spontaneous curvature [14,26,27], which cannot be measured by other MD methods. It is based on the calculation of the first moment of the stress profile, which gives the product of the bending modulus and spontaneous curvature. By dividing the latter by the bending modulus, one obtains spontaneous curvature. However, this approach usually relies on two methods decoupled from each other: the bending modulus is determined by some other method, not connected with the stress profile. Still, the stress profile allows determining the bending modulus itself. Namely, calculated at different stretching, the stress profiles make it possible to infer the local stretching modulus, $E(z)$, the second moment of which gives the bending rigidity [28], which is the consequence of the fact that bending is nothing but non-uniform stretching along the thickness of the monolayer. Moreover, once the local stretching modulus is known, it is possible to find other elastic parameters such as the stretching modulus and location of the neutral surface [6,17,28,29]. The stress profile approach also allows the investigation of lipid mixtures and potentially can be applied to other lipid systems such as monolayers in contact with water and oil [30] like in lipid droplets, bilayers in the gel phase [31], or membranes with adsorbed molecules [32]. Thus, this approach appears to be a self-consistent and widely applicable way of determining elastic parameters.

The local lateral stretching modulus acts as a characteristic function in membrane elasticity, as information contained in it allows one to calculate all substantial elastic parameters: the membrane stretching and bending moduli, spontaneous curvature, and location of the neutral surface. Previously, there were attempts to obtain $E(z)$ from simulations. In Refs. [33,34], $E(z)$ was calculated from the fluctuations of the depth-dependent area occupied by membrane components, without the requirement of any reference to the local stress. Within the local stress framework, till now, the only work where the local stretching modulus was calculated is Ref. [35]. However, the authors did not provide the error analysis and used simplified assumptions about the membrane compressibility, which led to the incorrect expression for $E(z)$ as we will show in the following.

In the current work, we present a theoretical framework and computational method for the measurement of elastic parameters of fluid bilayer membranes from the lateral stress profiles, that does not rely on the knowledge of the local stretching modulus, which largely simplifies the protocol. Unlike previous considerations of the membrane elasticity, we consider a more general condition of the global incompressibility, which is supported by experiments, instead of the local one. Considering the membrane at different ambient pressures, we establish a procedure for determining the local Poisson's ratio, which permits the determination of the exact scaling function that maps the points of the membrane at different stretching. This allows us to measure the deviation of the considered membranes from the incompressibility condition. The proposed method can be used not only for single-component systems but also for lipid mixtures. The underlying assumptions of our theory are based on the most general premises, from we which we obtain the expression for the local stress profile by performing the elastic energy variation. By analogy with the classical relation between the stretching modulus and lateral tension, we consider the relation between the bending modulus and corresponding stress moments and the Gaussian curvature modulus. We apply the developed method to Martini DOPC, DPPC, and their mixture to obtain the following elastic parameters of these systems: the monolayer bending and stretching moduli, spontaneous curvature, neutral surface location, and Poisson's ratio.

# Theory.

*Basic assumptions.* We consider the monolayers of a lipid membrane as a continuous elastic media. Our aim is to examine the forces arising in this medium due to deformations and relate them with the mechanical work necessary for stretching and bending, thus inferring the corresponding rigidities. We make the following widely accepted assumptions about each monolayer: i) lateral fluidity, ii) transverse isotropy, iii) global incompressibility, and iv) positivity of the isothermal local volumetric compressibility. The first two assumptions are the consequence of the fact that, when the membrane is flat, lipids can freely diffuse in the lateral directions, which leads to the isotropy of the monolayers in every plane lying parallel to the membrane, meaning that the monolayers are symmetric around the axis perpendicular to these planes. The third assumption follows both from experiments [36–39] and simulations [40], showing that the global bulk modulus of lipid membranes is close to that of water. Note that we distinguish the global incompressibility from the local one. The local incompressibility implies that every local volume element does not change during deformations. Although the local incompressibility implies the global one, the converse, in general, is not true, and MD simulations [40] show that the local compressibility may not hold. The fourth assumption merely demands the classicality of the material under consideration, as materials with negative compressibilities, showing the increase in volume in response to the increase in pressure, are rare and do not include lipid membranes [41].

Let's consider the elastic energy of the monolayers. We will analyze only deformations of stretching and bending whereby lipid molecules do not tilt. From the lateral fluidity and transverse isotropy, it follows that the monolayers have only two variables, which describe deformations: i) lateral stretching $\varepsilon$, i.e., the relative increase of the local area per lipid; and ii) the longitudinal stretching, $\varepsilon_\perp$, along the axis perpendicular to the plane of isotropy. Note however that since we also have the assumptions of the global incompressibility and positivity of the local bulk modulus, the variables $\varepsilon$ and $\varepsilon_\perp$ are not independent. Indeed, let's consider the monolayer under some uniform stretching $\varepsilon$. In this case, the thickness of the monolayer is completely determined by $\varepsilon$ due to the global incompressibility. Now, let's fix this thickness and start to apply higher pressure to the sides of the monolayer. Suppose that during this procedure longitudinal stretching, $\tilde{\varepsilon}_\perp$, counted relative to state before applying the higher pressure, becomes nonzero. This would imply that at some point $\tilde{\varepsilon}_\perp > 0$, as the total thickness is fixed. This, in turn, implies that at this point the local volume would increase in response to the increase in the external pressure, which however contradicts the assumption of the positivity of the bulk modulus. Therefore, $\varepsilon_\perp$ is a definite function at every $\varepsilon$. Note that this is true in uniform stretching, but might not be valid in bending. However, any deviations from the flat-state relation between $\varepsilon_\perp$ and $\varepsilon$, can be assumed to be small and neglected in accordance with the classical theory of elastic plates [42].

*Elastic energy and parameters.* Since $\varepsilon$ and $\varepsilon_\perp$ are not independent, the elastic energy density $W$ of the monolayer can be written as a function of only $\varepsilon$. We attribute this energy to the initial state, in which the monolayer is flat, and the total elastic energy equals to $\int_{V_0} W dV$, where $V_0$ is the initial volume occupied by the monolayer. Let's introduce a Cartesian coordinate system *xyz*, the *z*-axis of which coincides with the axis of symmetry and points from the hydrophilic to the hydrophobic part of the monolayer. Using this coordinate system, we can write the Taylor series of the elastic energy:

$$W(\varepsilon,z) = \sigma_0(z)\varepsilon(z) + \frac{1}{2}E(z)\varepsilon(z)^2 + ..., \qquad (1)$$

where the elastic parameters $\sigma_0$ and $E$ depend only on $z$ due to the lateral symmetry of the monolayer. In general, the local stretching $\varepsilon$ may also depend on $x$ and $y$, but, in the following, we will be interested only in the deformations with constant $\varepsilon$ along $x$- and $y$-axes. The function $E(z)$, being responsible for the quadratic contribution in $\varepsilon$, is the local stretching modulus, while $\sigma_0(z)$ corresponds to the pre-existing stress at $\varepsilon = 0$.

For each deformation, it is necessary to consider the compressibility condition:
$$\beta(\varepsilon,z) \equiv (1+\varepsilon(z))(1+\varepsilon_z(\varepsilon,z)), \tag{2}$$
where $\beta(\varepsilon)$, which we will call the *volume factor*, is the ratio of the deformed local volume to the initial one. Since at $\varepsilon \to 0$ $\beta(\varepsilon,z) \to 1$, we can employ the following results for the elastic constants, previously obtained within the incompressibility assumption in Refs. [6,16,29]:

$$k_A = \int_{m_0} E(z)\,dz, \tag{3a}$$

$$z_0 = \int_{m_0} E(z)\,z\,dz / k_A, \tag{3b}$$

$$k_m = \int_{m_0} E(z)(z-z_0)^2\,dz, \tag{3c}$$

$$K_{0,m} = -\int_{m_0} \sigma_0(z)\,z\,dz / k_m, \tag{3d}$$

where $\int_{m_0}$ denotes integration over the thickness of the monolayer in the initial state, $k_A$ is the area expansion modulus, $z_0$ is the location of the neutral surface, $k_m$ is the bending modulus relative to the neutral surface, and $K_{0,m}$ is spontaneous curvature. In general, the bending modulus over an arbitrary plane parallel to the monolayer is given by the expression $\tilde{k}_m = \int E(z)(z-\tilde{z})^2\,dz$, where $\tilde{z}$ is the coordinate of the chosen plane. There however exists such a plane, called the neutral plane or neutral surface, where the bending modulus is minimal and the location of which, given in Eq. (3b), follows from the equation $\partial \tilde{k}_m / \partial \tilde{z} = 0$. The bending modulus relative to this plane is of the highest interest, as it represents the easiest way to bend the monolayer.

## Theoretical results.

*Force factor.* In this section, we provide the theoretical basis for the method of the measurement of the elastic parameters of lipid membranes. These parameters are given in Eqs. (3a–d) as the *n*-th moments of $E(z)$. Being the local stretching modulus, $E(z)$ should be the derivative with respect to $\varepsilon$ at $\varepsilon = 0$ of $F(\varepsilon,z) \equiv \dfrac{\partial W(\varepsilon,z)}{\partial \varepsilon}$, which we will call the *force factor*. Inserting the relation $E(z) = \dfrac{\partial F(\varepsilon,z)}{\partial \varepsilon}\bigg|_{\varepsilon=0}$ into Eqs. (3a–d) and interchanging the derivative and integral signs, we obtain:

$$k_A = \frac{d}{d\varepsilon}\int_{m_0} F(\varepsilon,z)\,dz \bigg|_{\varepsilon=0} \tag{4a}$$

$$z_0 = \frac{d}{d\varepsilon}\int_{m_0} F(\varepsilon,z)\,z\,dz \bigg|_{\varepsilon=0} / k_A \tag{4b}$$

$$k_m = \frac{d}{d\varepsilon}\int_{m_0} F(\varepsilon,z)(z-z_0)^2\,dz \bigg|_{\varepsilon=0} \tag{4c}$$

$$K_{0,m} = -\int_{m_0} F(0,z)\, z\, dz / k_m. \tag{4d}$$

$F(\varepsilon, z)$ can be found by the variation of the elastic energy with respect to $\varepsilon$ (see Appendix A):

$$F(\varepsilon, z) = \frac{(S(\varepsilon, z, P_z) + P_z)\beta(\varepsilon, z)}{1+\varepsilon} - P_z \frac{\partial \beta(\varepsilon, z)}{\partial \varepsilon}, \tag{5}$$

where $P_z$ is the external pressure, i.e., 1 bar under normal conditions; $S(\varepsilon, z, P_z)$ is the lateral stress. The sign convention we use in Eq. (5) is such that negative $S(\varepsilon, z)$ means repulsion.

*Scaling.* The stress profile $S(\varepsilon, z, P_z)$ in Eq. (5) is given relative to the reference state at $\varepsilon = 0$, i.e., $S(\varepsilon, z(\zeta), P_z) \equiv \tilde{S}(\varepsilon, \zeta, P_z)$, where $\tilde{S}(\varepsilon, \zeta, P_z)$ and $\zeta$ are the lateral stress and coordinate along the axis of symmetry in the current state of the monolayer, respectively. The relation between $z$ and $\zeta$ can be described by some scaling function $z(\zeta)$, which maps the points in the reference state to the corresponding points in the stretched state of the monolayer. $z$ and $\zeta$ are not equal since the membrane thickness changes under uniform stretching. In the case of the local incompressibility, the relation between $z$ and $\zeta$ would be $\zeta = (1+\varepsilon)^{-1} z$, which we will call the *uniform scaling*. In the more general case of the global incompressibility, the scaling function can be found from the compressibility condition, $(1+\varepsilon)\nabla_z \zeta(\varepsilon, z) = \beta(\varepsilon, z)$:

$$\zeta(\varepsilon, z) = (1+\varepsilon)^{-1} \int_0^z \beta(\varepsilon, t)\, dt, \tag{6}$$

which we will refer to as the *exact scaling*. To determine $\beta(\varepsilon, z)$, we can take the derivative of Eq. (5) with respect to $P_z$ at $\varepsilon = 0$ and define:

$$\gamma(z) \equiv \left.\frac{\partial \beta(\varepsilon, z)}{\partial \varepsilon}\right|_{\varepsilon=0} = \frac{\partial S(0, z, P_z)}{\partial P_z} + 1. \tag{7}$$

Thus, the derivative of $\beta(\varepsilon, z)$ is determined by the response of the lateral stress profile to the change in the ambient pressure. Direct calculations from MD data show that $\beta(\varepsilon, z)$ can be approximated linearly: $\beta(\varepsilon, z) = 1 + \gamma(z)\varepsilon$, and higher-order terms in $\varepsilon$ are negligibly small.

*Poisson's ratio.* $\gamma(z)$ also allows the determination of the local Poisson's ratio, an important parameter that describes the deviation of material from the local incompressibility. It is defined as $\nu(z) \equiv \frac{1}{2}\lim_{\varepsilon \to 0} -\frac{\varepsilon}{\varepsilon_z(z)}$ and equals 0.5 in the case of the local incompressibility. Inserting $\varepsilon_z(z)$ from Eq. (2), into the definition of $\nu(z)$, we get:

$$\nu(z) = \frac{1}{2(1-\gamma(z))}. \tag{8}$$

Thus, $\gamma(z)$, given in Eq. (7), which can be directly measured in MD simulations, provides the information both about the local volume factor and the local Poisson's ratio. From Eq. (8) it follows that in the case of the local incompressibility $\gamma(z) = 0$, which implies, according to Eq. (7), that additional ambient pressure evenly spreads over the lateral stress profile. In general, however, $\gamma(z)$ might be different from zero, which would imply the deviation from the local incompressibility.

*Integral properties.* Let us now consider what integral properties follow from Eqs. (4a) and (4c) for the stretching and bending moduli. It is convenient to introduce a force factor relative to the state at stretching $\varepsilon$: $F_\varepsilon(\tilde{\varepsilon}, \zeta) \equiv \frac{\partial \tilde{W}(\tilde{\varepsilon}, \zeta)}{\partial \tilde{\varepsilon}}$, where $\tilde{\varepsilon}$ and $\tilde{W}(\tilde{\varepsilon}, \zeta) \equiv \frac{W(\varepsilon(\tilde{\varepsilon}), z(\zeta))}{\beta(\varepsilon)}$ is stretching and the elastic energy density written relative to this state. Then, using the equality $F(\varepsilon,z)dz = F_\varepsilon(0,\zeta)d\zeta$, from Eq. (4a) we get the classical expression $k_A = \dfrac{d}{d\varepsilon}\sigma(\varepsilon)\Big|_{\varepsilon=0}$, where $\sigma(\varepsilon) \equiv \int_{m_\varepsilon} F_\varepsilon(0,\zeta)d\zeta$ is by definition the lateral tension at stretching $\varepsilon$. If we consider Eq. (4c) for the bending modulus, it can be rewritten as:

$$k_m = \frac{d}{d\varepsilon}\int_{m_\varepsilon} F_\varepsilon(0,\zeta)(z(\zeta)-z_0)^2 \, d\zeta \Big|_{\varepsilon=0}. \tag{9}$$

The integral on the right-hand side resembles the Gaussian curvature modulus [6,43,44] measured at finite stretching $\varepsilon$, $\bar{k}_m(\varepsilon) = \int_{m_\varepsilon} F_\varepsilon(0,\zeta)(\zeta-\zeta_0)^2 \, d\zeta$. The difference is that the distance from the neutral surface is taken relative to the reference state at $\varepsilon = 0$: $z(\zeta)-z_0$ instead of $\zeta-\zeta_0$, where $\zeta_0$ is the neutral surface location at stretching $\varepsilon$. In the case of the local incompressibility and quadratic energy law, $z(\zeta)-z_0 = (1+\varepsilon)(\zeta-\zeta_0)$, which turns Eq. (9) into:

$$k_m = 2\bar{k}_m(0) + \frac{d}{d\varepsilon}\bar{k}_m(\varepsilon)\Big|_{\varepsilon=0}, \tag{10}$$

Thus, the dependence of the Gaussian curvature modulus on stretching contains information about the bending rigidity. Below, among other things, we will test the validity of Eq. (10) and its assumptions in MD simulations.

## Methods

To obtain stress profiles, we performed a coarse-grained MD simulation. We used the Martini 3 [45] force field to represent the lipid bilayers. Each system consisted of 256 lipids and 11 water beads per lipid. MD simulations were run with GROMACS 2020 [46,47] program package. The long-range electrostatics was treated with the reaction field method and relative dielectric constant ε = 15. The cutoff for the Lennard-Jones and Coulomb interactions was set to 1.2 nm. The temperature at 300 K was controlled with the velocity-rescale thermostat [48]. For the solvent and bilayers, two separate temperature coupling groups were introduced, and the coupling constant was set to $\tau_T$ = 1 ps. To control the constant surface tension the Berendsen barostat [49] with the surface-tension coupling was employed with the time constant $t_P$ = 3 ps and isothermal compressibility $k_T$ = 3 x 10$^{-5}$ Pa. For all simulations, the 20 fs time step was used. The systems were built and equilibrated using the CHARMM-GUI web server [50–52].

*Resampling stress and grid data.* Every system at each tension was simulated for 1000 ns after equilibration for 100 ns. The MD simulations were performed at surface tensions of −100, −50, 0, 75, and 150 bar·nm. For the statistical analysis, we divided the simulation time into 20 blocks of 50 ns each. Each block was then analyzed with GROMACS-LS [53–56] whereby the covariant central force decomposition was employed and the lateral stress was determined as an average between *xx*- and *yy*- components of the obtained local stress. Before the stress analysis, the simulation box was centered with respect to the membrane center of mass. The grid step for the local stress was set to 0.05 nm. One additional point was added after the last grid point to make the lateral stress periodic. Then, each stress profile was centered by subtracting the coordinate of the last grid point from grid coordinates and dividing by two. Thus, the mid-plane of the membrane occurred at $z = 0$. After that, the stress profile of each block was symmetrized by adding the stress profile in reverse order to the initial stress profile and dividing by two. Then, from the 20 obtained stress profiles, 20 stress profiles were randomly sampled with repetition and averaged between each other. The latter was repeated 200 times, and thus 200 stress profiles were obtained. The grid data in GROMACS-LS represents the average grid over the simulation time. Therefore, along with 200 profiles, 200 grids were obtained by averaging the corresponding grid data. To determine the stretching $\varepsilon$, at each tension $\sigma$ the lateral size of the simulation box, $L(\sigma)$, was determined as $L(\sigma) = (L_x(\sigma) + L_y(\sigma))/2$, where $L_x(\sigma)$ and $L_y(\sigma)$ are the average box lengths along the *x*- and *y*-axes over the entire simulation time of 1000 ns. Then, the stretching was determined as $\varepsilon = L(\sigma)^2 / L(0)^2 - 1$.

*Scaling.* In the case of the uniform scaling, all 200 grids of the stress profiles were divided by $1 + \varepsilon_z(\sigma)$, where $\varepsilon_z(\sigma) = L_z(\sigma)/L_z(0) - 1$ with $L_z(\sigma)$ being the longitudinal size of the box which was determined as the average size of the 200 obtained grids.

The exact scaling, given by Eq. (6), requires the calculation of $\beta(\varepsilon, z)$. To do that, the linear approximation, $\beta(\varepsilon, z) = 1 + \gamma(z)\varepsilon$, was employed, as it can be shown that the second derivative $\left.\frac{\partial^2 \beta(\varepsilon, z)}{\partial \varepsilon^2}\right|_{\varepsilon=0}$ is negligibly small (see Appendix B). To determine $\gamma(z)$, given in Eq. (7), it is necessary to apply the isotropic pressure to the membrane. To achieve this, we set the surface tension to zero and varied $P_z$, setting the following pressures: −50, −25, 1, 25, and 50 bar. At each $P_z$, the simulations were performed with the same protocol as described above. We note that despite the global incompressibility assumption the longitudinal size of the box may slightly change as the isotropic pressure is applied. This change, however, is rather small and in the case of, for instance, DPPC membrane does not exceed 0.2%. Nevertheless, to make the grid sizes match each other, we perform the uniform scaling in the same way as described above for the cases of nonzero tensions. Then, $\frac{\partial S(0, z, P_z)}{\partial P_z} + 1$ was found by a linear least-squares estimator. The latter was performed 200 times consequently for all 200 stress profiles at different $P_z$. Thus, 200 random values of $\gamma(z)$ were obtained. Then, the Poisson's ratio and the exact scaling map were found by inserting the values of $\gamma(z)$ and $\beta(\varepsilon, z) = 1 + \gamma(z)\varepsilon$ into Eqs. (8) and (6), respectively. After that, the obtained scaling maps were applied to grid data and 200 force factors at each stretching were determined via Eq. (5).

*Elastic parameters.* Once the force factor, $F(\varepsilon, z)$, is determined, we can use Eqs. (4a–d) to find the elastic constants, calculating the moments of $F(\varepsilon, z)$. For this, at each $\varepsilon$ one random force factor of the 200 available was chosen. Then, the stretching modulus, location of the neutral surface, bending modulus, and spontaneous curvature were determined consequently according to Eqs. (4a–d). To find the derivatives with respect to $\varepsilon$, quadratic least-squares regression was performed. This procedure was repeated multiple times to get sufficient sampling statistics.

## Simulation results

In this section, we present the results of the MD simulations. We start with DPPC, a widely used lipid in the simulations. Firstly, we find the Poisson's ratio $\nu(z)$ (Figure 1), calculating the response $\gamma(z)$ of the lateral stress profile to the external isotropic pressure $P_z$ according to Eq. (7) and inserting it into Eq. (8). The following isotropic pressures were applied: −100, −50, 1, 50, and 100 bar. A linear least-squares fit is used to find $\gamma(z)$, as at each *z* a linear function agrees well with the data within the error limits.

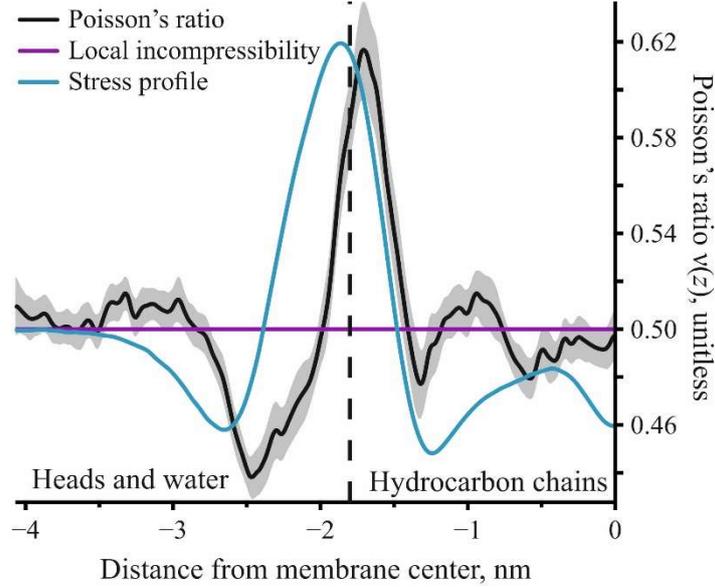

Figure 1. The Poisson's ratio of DPPC at 325 K as a function of the distance from the membrane center. The black curve and gray color indicate the mean value of the Poisson's ratio and 95% confidence band, respectively. The purple line shows the local incompressibility case, wherein the Poisson's ratio equals 0.5. The zero-tension stress profile, adjusted to the magnitude of the Poisson's ratio, is given in arbitrary units by the cyan curve. The dash vertical line shows the mean coordinate of the first glycerol bead, ≈ 1.8 nm, indicating the boundary between the hydrophobic and hydrophilic parts of the monolayer.

Figure 1 shows that at a lot of coordinate points the Poisson's ratio differs from 0.5, i.e., from the incompressibility condition. We observe a correlation between the stress profile and Poisson's ratio: in the head group region, where the stress is repulsive, the Poisson's ratio is less than 0.5, meaning that the magnitude of the longitudinal stretching is larger than that of the lateral one during deformations. By contrast, in the glycerol region ($z \approx -1.8$ nm), where the attraction occurs, the ratio is greater than 0.5 and reaches the maximum deviation from 0.5: $\nu = 0.62 \pm 0.01$. In the repulsive head group region, the Poisson's ratio is again smaller than 0.5.

After the Poisson's ratio is found, we can determine the volume factor $\beta(\varepsilon,z) \equiv 1 + \gamma(z)\varepsilon$ and the exact scaling function, Eq. (10). We employ the linear approximation of $\beta(\varepsilon,z)$ because our direct calculations show that the difference between the Taylor series of $\beta(\varepsilon,z)$ up to the first and second order in $\varepsilon$ is less than $\approx 0.3\%$ within the considered range of $\varepsilon$. The moments of $\left.\frac{\partial^2 \beta(\varepsilon,z)}{\partial \varepsilon^2}\right|_{\varepsilon=0}$ are also negligibly small; the details of the calculations are given in Appendix B. This allows us to use the linear approximation for $\beta(\varepsilon,z)$ in the expression for the force factor, Eq. (5).

To obtain the values of the force factor at various stretching, we apply the following tensions: −100, −50, 75, and 150 bar·nm. The absolute values of the negative tensions are chosen to be smaller, as high values might lead to a large compression of the membrane with the formation of a gel phase. The latter may occur at some critical value of compression, depending on a lipid system and temperature. DLPC at 300K, for example, transitions into a gel phase at ≈ 3.8% [57]. In our simulations, the largest compression was 3.2% and we did not observe any discontinuities in the obtained data.

The stress profiles at different tensions have different endpoints, as the membrane shrinks along the lipid tails in response to stretching. Using $\beta(\varepsilon,z)$, we can find the exact scaling function, given in Eq. (6), to map the points of the stretched membranes to the points of the initial state at zero tension.

Applying this scaling map to the stress profiles, we determine the force factors at different stretching and then find the elastic parameters as given in Eqs. (4a–c), performing the least-squares fit to the moments of the force factors at different $\varepsilon$. The results are presented in Figure 2.

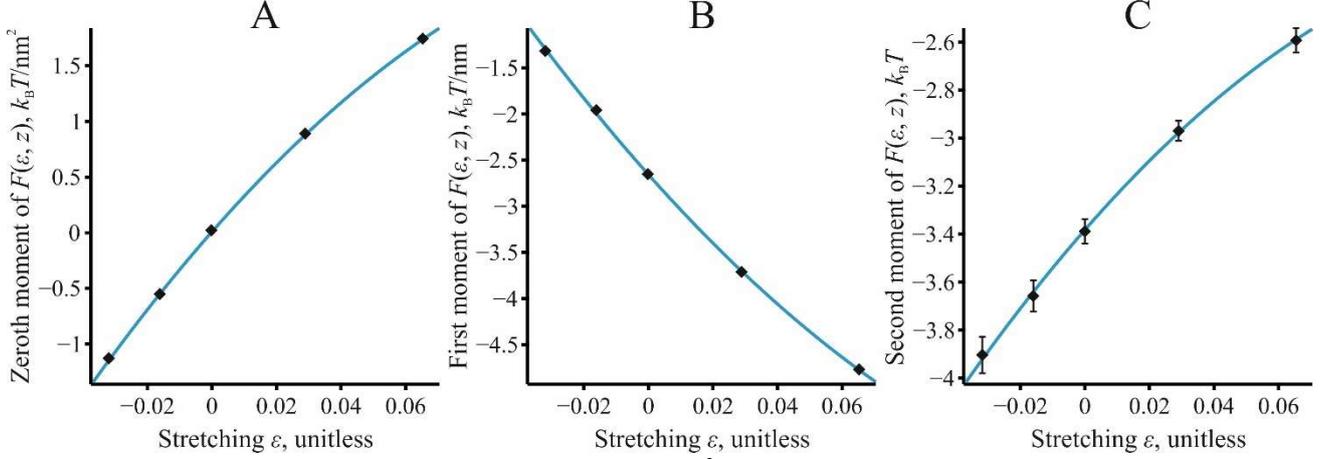

Figure 2. (A) The zeroth moment of the force factor, $\int_{m_0} F(\varepsilon, z)\, dz$, (B) the first moment of the force factor, $\int_{m_0} F(\varepsilon, z)\, z\, dz$, and (C) the second moment of the force factor, $\int_{m_0} F(\varepsilon, z)(z - z_0)^2\, dz$, where $z_0$ is the location of the neutral surface, as functions of stretching $\varepsilon$. The cyan curves are the best quadratic fits to the data. The error bars in panel C indicate the standard deviation. In panels A and B the error bars are too small to be shown.

For each data, to find the derivative at $\varepsilon = 0$ we use the quadratic least-squares estimator, as fitting with higher-order polynomials does not change the mean value within the error limits. We find the following values for the elastic parameters: $k_A = 33.0 \pm 0.2\ k_B T/\mathrm{nm}$, $z_0 = -1.19 \pm 0.01$ nm, $K_{0,m} = -0.17 \pm 0.01$ nm$^{-1}$, and $k_m = 15.3 \pm 0.6\ k_B T$ for the monolayer stretching modulus, location of the neutral surface, spontaneous curvature and bending modulus, respectively. The errors indicate the standard deviation. As seen in Figure 2, the quadratic fit well describes the data; the reduced chi-squared is 2.23, 1.86, and 1.02 for the zeroth, first and second moment, respectively. Note that, since $F(0, z)$ coincides with the tension profile at zero stretching, the values at $\varepsilon = 0$ in Figure 2B and Figure 2C are equal to $-k_m K_{0,m}$ and the Gaussian curvature modulus, respectively.

Next, we proceed by changing the scaling map from the exact to the uniform one to find out the degree of the corrections to the elastic parameters resulting from the deviation from the incompressibility. The results are $k_A = 32.1 \pm 0.1\ k_B T/\mathrm{nm}^2$, $z_0 = -1.16 \pm 0.01$ nm, $K_{0,m} = -0.18 \pm 0.01$ nm$^{-1}$, and $k_m = 14.8 \pm 0.6\ k_B T$ for the stretching modulus, location of the neutral surface, and bending modulus, respectively. There is a slight shift in the mean values in comparison with the exact scaling. Although the discrepancy does not fall within the error limits for the stretching modulus and neutral surface, the relative deviation is less than 3%. The same observation has been made for other lipid systems (DOPC and DPPC/DOPC mixture) and temperatures. We thus may conclude that the choice of the scaling method does not noticeably influence the results of calculations, i.e. the deviation from the incompressibility does not contribute much to the values of the considered elastic constants. Therefore, we recommend employing the easier and more trackable uniform scaling.

Employing the same method, we next consider the 50:50 DPPC/DOPC lipid mixture and pure DOPC at 325 K. The results are presented in Figure 3 and Table C1. The stretching and bending moduli demonstrate opposite trends as functions of the unsaturated lipid concentration: while the stretching constant and spontaneous curvature increase with the DOPC concentration, the bending rigidity decreases. Neutral surface coordinate shifts towards the mid-plane of the membrane and vary almost linearly at the considered concentrations of DOPC. The variation in the bending modulus and spontaneous curvature however is highly nonlinear: the difference between pure DPPC and the mixture is much larger than the difference between the mixture and pure DOPC. As should be expected, the bending modulus of DOPC turns out to be smaller than that of DPPC since DOPC is an unsaturated lipid, unlike DPPC. The maximum values of the deviation of the Poisson's ratio from the incompressibility condition $|\nu(z) - 0.5|_{\text{Max}}$ decrease from $0.12 \pm 0.01$ to $0.07 \pm 0.01$.

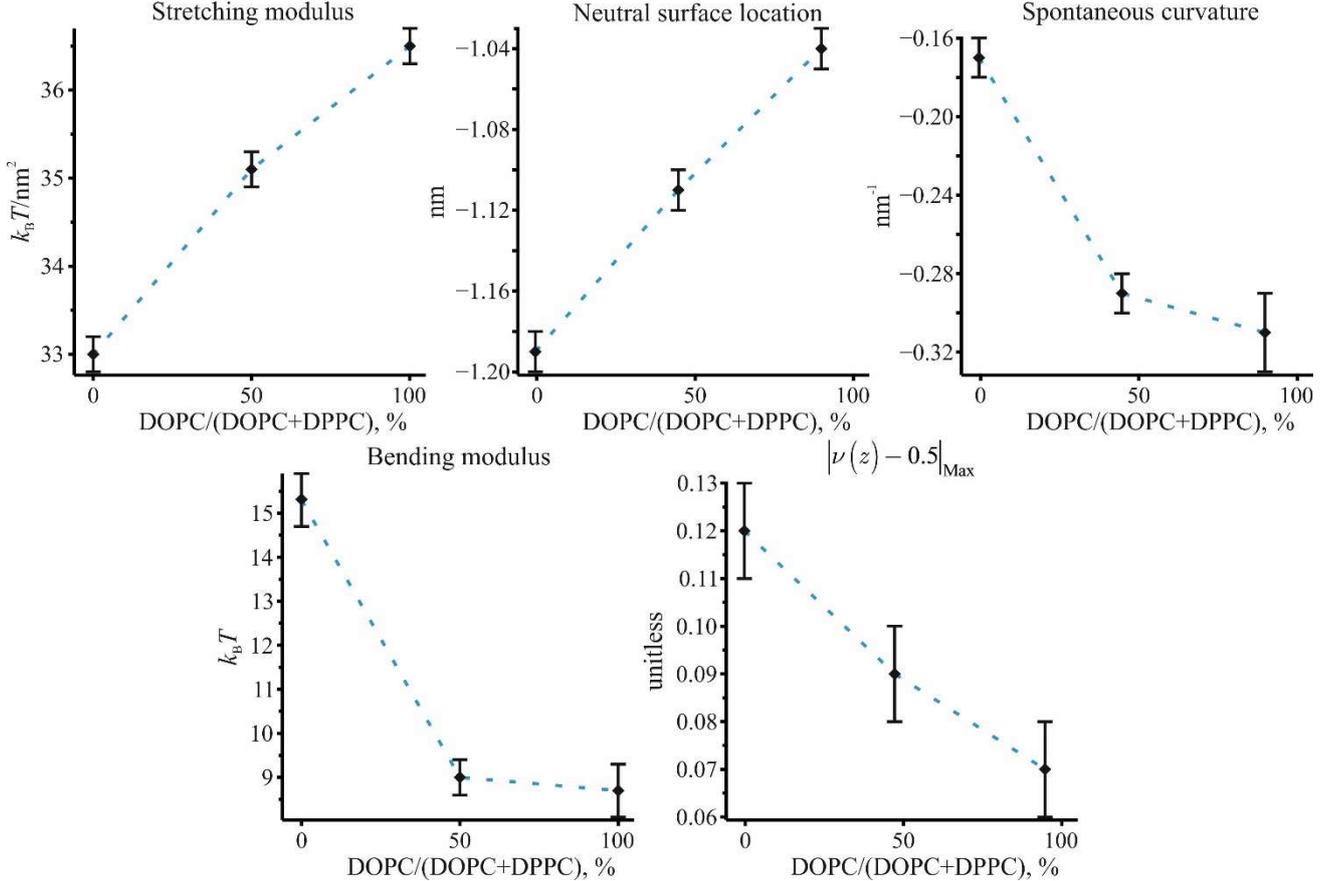

Figure 3. The monolayer elastic parameters as functions of DOPC concentration in the DPPC/DOPC mixture, measured using Eqs. (4a–d) and (8). The error bars indicate the standard deviation.

We also considered DOPC at 300 K to capture the dependence of the elastic constants on temperature and to see whether the proposed method also works at smaller temperatures (Table I). With the decrease in temperature, the membrane becomes more rigid: the stretching and bending moduli become larger. This is anticipated, as the membrane gets closer to the phase transition into a gel phase with the decrease in temperature. At the same time, the location of the neutral surface, spontaneous curvature and $|\nu - 0.5|_{\text{Max}}$ remains approximately the same.

Eq. (10) relates the bending and Gaussian curvature moduli under the assumptions of the quadratic energy and locally incompressible membrane. The direct calculations for DPPC yield that the Gaussian curvature modulus equals $-3.40 \pm 0.05\ k_BT$ and its derivative is $14.8 \pm 2.2\ k_BT$, leading to $8.1 \pm 2.2\ k_BT$ for the bending rigidity, which is far from the value of $15.3 \pm 0.6\ k_BT$ obtained from Eq. (4c). This implies that the assumptions of the quadratic energy law and local incompressibility are not valid, which agrees with the previous observation of the local Poisson's ratio deviating from the local incompressibility condition. The energy deviation from the quadratic law can also be seen from the dependence of the monolayer tension on stretching, which is not linear (see Figure 2A) as it would be in the quadratic energy model. Thus, the quadratic incompressible model appears to be oversimplified, at least for the coarse-grained DPPC.

## Discussion

In this paper, we proposed the theoretical framework and method for the measurement of elastic parameters of lipid membranes, based on the lateral stress profiles at different stretching $\varepsilon$ obtained by the means of the MD simulation. Introducing the notion of the force factor, we demonstrated how to extract the elastic parameters from fitting the force factor moments at different stretching. In addition, the introduced notion of the volume factor allowed us to find the Poisson's ratio that characterizes the deviation of the membrane from the incompressibility condition and determines the exact scaling map between the stress profiles of the stretched states of the membrane. We applied the method to the coarse-grained DPPC, DOPC, and DPPC/DOPC mixture at different temperatures.

To justify the method, we considered the membrane as a 3D elastic body and obtained the expression for the stress tensor, Eq. (A5). Previously lipid membranes were modeled as either incompressible [6,14,16,29,58] or as a material with independent lateral and longitudinal stretching [40]. In our work, we instead employed the global incompressibility assumption, supported by experiments [36–39], and showed that lateral and longitudinal stretching are not independent. This allowed us to express the energy as being dependent only on lateral stretching and obtain the expression, given in Eq. (8), for the local Poisson's ratio that characterizes the deviation from the local incompressibility. Note that within the proposed framework the Poisson's ratio is fully determined by the lateral stress profiles, i.e., without any reference to the amount of material under consideration. It is however possible to determine the Poisson's ratio from the material perspective, as was made by *Terzi et al.* [40]. An interesting question therefore is how the two approaches are related to each other. Definitely, both should lead to the same Poisson's ratio. While the *Terzi et al.*'s approach depends on the definition of the amount of material, which may contain some uncertainties with, for example, additivity of component volumes, our procedure relies only on the lateral stress profile. Still, the definition of the latter may be complicated since it largely depends on the force decomposition used [53]. It is argued though that some decompositions are preferable over others. The Goetz-Lipowsky decomposition, for example, does not conserve the angular momentum of the system as opposed to the central force decomposition [53], which was used in this paper. However, in Ref. [59] it was shown that the central force decomposition is not the only one that conserves the angular momentum. Thus, we see that there is currently an ambiguity in the stress tensor determination methods in the literature. To circumvent this uncertainty in the definition of the stress profiles, a method has been proposed that analytically calculates the elastic parameters of lipid membranes from the given local interacting potentials [60]. The latter approach, however, still depends on the choice of the center of the in-plane deformations.

For the determination of the elastic parameters, we used the notion of the force factor $F(\varepsilon,z) \equiv \dfrac{\partial W(\varepsilon,z)}{\partial \varepsilon}$, introduced in Eq. (5), where $W(\varepsilon)$ is the elastic energy density. By definition, $F(\varepsilon,z)$ represents the resulting effective lateral stress given in the reference state, which does work during deformations. In this regard, $\int_{m_0} F(\varepsilon,z)\,dz$ gives by definition the lateral tension at the state of stretching $\varepsilon$. The relations between elastic constants and $F(\varepsilon,z)$ are given in Eqs. (4a–d). The bending modulus, for example, can be found as a derivative with respect to $\varepsilon$ at $\varepsilon = 0$ of the second moment of $F(\varepsilon,z)$. It should be noted that there is also another way to determine elastic parameters.

The force factor is directly related to the local stretching modulus $E(z)$, which is the derivative of $F(\varepsilon, z)$ at $\varepsilon = 0$. Thus, $E(z)$ could have been used, according to Eqs. (3a–d), without the use of $F(\varepsilon, z)$. It would require however the determination of $E(z)$ at some specified grid distributed throughout the thickness of the monolayer. The finer the grid, the better the result for the moments of $E(z)$ would be. As the grid step should be at least as small as the grid used in the calculation of the stress profile, which is 0.05 nm in our case, the number of grid points for the monolayer with 4 nm in thickness would be 80. This largely complicates the procedure and statistical analysis since at each point of the grid it is necessary to perform fitting of the data to determine the derivative of the force factor. We showed though that this is a superfluous task, as the same goal can be achieved with only one fit of the force factor moments. Actually, $E(z)$ requires too much information to be calculated: once $E(z)$ is determined, any moment of it can be found. This information, however, is redundant for the determination of the key elastic parameters such as stretching and bending moduli and spontaneous curvature. Previously, the local stretching modulus profile, in the case of Martini DOPC, was found in Ref. [35] wherein $E(z)$ was equated with the derivative of the stress profile with respect to stretching. However, as follows from Eq. (5), even in the incompressible case, the expression for $E(z)$ is more complicated: it contains not only the derivative of the stress profile but also the stress itself, which is comparable in magnitude with its derivative. This is the consequence of the change of the membrane thickness due to the applied lateral tension, which may lead to incorrect results if not taken into account. For instance, the bending modulus of DPPC would be 11.3 ± 0.6 $k_B T$ instead of 14.8 ± 0.6 $k_B T$ in the case of the uniform scaling.

For all considered systems, we have calculated the local Poisson's ratio, $\nu(z)$, which quantifies the deviation from the local incompressibility. The local incompressibility implies the constant Poisson's ratio equal to 0.5. Our calculations showed that the Poisson's ratio of the considered systems is inhomogeneous: it may be smaller or larger than 0.5, depending on the considered region (see Figure 1). The maximum values of $|\nu(z) - 0.5|$ lie within the range of 0.57–0.62, i.e., 14–24% away from 0.5. Such a deviation from the incompressibility leads to the discrepancy between the values of elastic parameters obtained under the uniform and exact scaling. This discrepancy however is not large: for the stretching modulus, for example, it is less than 3%, albeit larger than the error limits.

In this work, the coarse-grained models were used. Yet, the proposed method applies to any lipid system which allows the calculation of the stress profile, including AA ones. The coarse-grained model was chosen not only because it is fast and convenient for testing the proposed method, but also because the current version of GROMACS-LS does include the particle-mesh Ewald method for the electrostatic calculation needed for AA simulations, without which the calculation requires a high computational cost [53].

As the coarse-grained systems were considered in this paper, we should not anticipate a full coincidence of the obtained elastic parameters with experimental results. In turn, different experimental methods may also disagree with each other. The analysis of thermal shape fluctuations of vesicles at 322 K, for example, leads to 18.4 ± 1.1 $k_B T$ for the monolayer bending modulus of DPPC [61], while the X-ray scattering at 323 K yields 14.4 ± 2.3 $k_B T$ [62]. Nevertheless, the value obtained in this work, 15.3 ± 0.6 $k_B T$ at 325 K, is quite close to these experimental results. The same is also true for the obtained values of the bending modulus of DOPC, 8.7 ± 0.6 $k_B T$ at 325 K and 10.8 ± 0.8 at 300 K, and corresponding experimental results (10.6 ± 1.2 $k_B T$ at 291 K [63], 7.5 ± 1.5 $k_B T$ at 295 K [64], 9.7 ± 0.4 at 303 K [62]). The values of the stretching moduli are also typical [65,66]. The neutral surface of the considered systems lies ≈ 1 nm away from the mid-plane, i.e., approximately in the middle of the tails region. This is in agreement with previous simulations based on the analysis of fluctuations [14]. What is noticeable is that the obtained values of spontaneous curvature are large and negative (see Table C1). In contrast, the experimental values are 0.05 ± 0.05 nm$^{-1}$ and −0.04 ± 0.04 nm$^{-1}$ for DPPC and DOPC, respectively [67]. This might be an artifact of the Martini force field which leads to the large repulsion in the tails region (see Figure 1), whereas in AA systems this repulsion is smaller [53]. The consequence of this may be the increase in the line tension of the membrane edge and therefore the high cost of the pore formation, which is a known drawback of the Martini lipids [68].

Since the proposed method can be used not only for pure lipids but also for lipid mixtures, we can observe a concentration trend for spontaneous curvature along with other elastic parameters. It should be noted, however, that in general the bending deformation might entail the redistribution of the membrane components due to the composition-curvature coupling [69], leading to the effective bending modulus, which is smaller than that without the redistribution effect. The method, proposed in the present work, considers only uniform systems and does not take into account the redistribution effect. Therefore, the bending moduli obtained for mixtures represent the upper bound for the effective bending modulus. The decrease in the bending rigidity due to the redistribution of lipids might be up to ~30%, and theoretical models exist that estimate this decrease from the properties of the reservoir and the dependence of the bending modulus and spontaneous curvature of uniform membranes on the concentrations of components [70]. For this work, we considered the 50:50 DPPC/DOPC lipid mixture. While the stretching modulus of this mixture turned out to be somewhere in the middle between those of pure DPPC and DOPC, the obtained values of the bending modulus and spontaneous curvature are close to pure DOPC. Thus, the trend for the latter two parameters is highly nonlinear. The inverse of the bending modulus also does not abide by a linear trend, known as Reuss averaging [71], sometimes used in simplified theoretical models [72,73]. According to this averaging, the bending modulus of a binary mixture from lipids with the bending moduli $k_1$, $k_2$ and areas per lipid $a_1$, $a_2$ of the corresponding single-component membranes equals $(a_1 N_1 + a_2 N_2)\left(\dfrac{a_1 N_1}{k_1} + \dfrac{a_2 N_2}{k_2}\right)^{-1}$, where $N_1$, $N_2$ are the number of lipids in the mixture. Substituting here, the areas per lipid of the pure DPPC and pure DOPC membranes, 0.6 nm$^2$ and 0.69 nm$^2$, respectively, and the obtained values of the bending moduli, we get 10.9 ± 0.5 $k_B T$ for the bending modulus of the 50:50 DPPC/DOPC mixture, which differs from the measured value, 9.0 ± 0.4 $k_B T$.

In real systems and simulations, lipid membranes undergo fluctuations due to thermal motion, which gives rise to out-of-plane undulations. The excess area, absorbed by these undulations, may lead to corrections to the observed values of forces and apparent stretching [66,74]. This is particularly important for large systems since the excess area contribution is also large. However, in small systems the corrections are minor. For example, for systems as large as those employed for buckling, ≈ 45 × 7 nm, the corrections do not exceed 1% [22]. In our case, the considered systems are even smaller with the lateral size of the box being ≈ 10 nm. Therefore, we did not include the effect of thermal fluctuations in the analysis of the elastic parameters.

The obtained expression for the force factor, Eq. (5), allowed us to consider the relation between the bending modulus and the moments of the stress profiles. This analysis is analogous to the derivation of the classical relation between the stretching modulus and lateral tension: $k_A = \dfrac{d}{d\varepsilon}\sigma(\varepsilon)\big|_{\varepsilon=0}$. We showed that in the case of the quadratic energy law and local incompressibility the bending modulus is associated with the $\varepsilon$-dependent Gaussian curvature modulus according to Eq. (10). This is a consequence of the fact that the expression for the bending modulus, Eq. (4c), contains the second moment of the stress profile. It might seem striking that the two elastic moduli, the bending modulus and the Gaussian curvature modulus, at least in the case of the quadratic energy law and local incompressibility, are not independent of each other. However, it was previously shown that the second moment of the stress profile, which is usually associated with the Gaussian curvature modulus [6,43,44], should not be considered as a separate elastic modulus: rather, this term is connected with spontaneous Gaussian curvature [29]. Also, the reformulation of the Gaussian curvature energy term in terms of the deviatoric bending and spontaneous warp was suggested in Ref. [75]. Our analysis showed that Eq. (10) does not hold for DPPC: $k_c$ calculated from Eq. (10), $8.1 \pm 2.2$ $k_BT$, is approximately half the actual value, $15.3 \pm 0.6$ $k_BT$. This implies that the simplifying assumption of the local incompressibility and quadratic energy law does not apply to Martini DPPC.

## APPENDIX
### A. Variation of the elastic energy

In this appendix, we derive the expression for the force factor, Eq. (5). The elastic energy function of the monolayer is given by $W(\varepsilon, z)$. To perform the variation with respect to $\varepsilon$, we need to provide the geometrical structure this energy function describes. To do this, we follow the approach suggested in Ref. [76], wherein the geometry is specified with the help of Lagrange multipliers. Let's firstly introduce the coordinate system $(x_1, x_2, x_3)$, the $x_3$-axis of which is directed along the lipid tails from the hydrophilic to the hydrophobic part of the monolayer, and $x_1$, $x_2$ describe the points of the transverse planes at a given value of $x_3$. Using this coordinate system, we can describe the points of the initial and deformed states of the monolayer with the help of the vector functions $\mathbf{R}_0(x_1, x_2, x_3)$ and $\mathbf{R}(x_1, x_2, x_3)$, respectively. These functions give rise to the local bases and the metric tensors via $\mathbf{e}_{0a} \equiv \partial_a \mathbf{R}_0$, $\mathbf{e}_a \equiv \partial_a \mathbf{R}$, $g_{0ab} \equiv \mathbf{e}_{0a} \cdot \mathbf{e}_{0b}$, $g_{ab} \equiv \mathbf{e}_a \cdot \mathbf{e}_b$, $g_0 \equiv \det g_{0ab}$, $g \equiv \det g_{ab}$, $a, b = 1, 2, 3$, and the area densities of the transverse planes $x_1, x_2$: $\hat{g} = \det g_{ab}$, $\hat{g}_0 = \det g_{0ab}$, $a, b = 1, 2$. Now, we can construct the following Hamiltonian:

$$\begin{aligned}
H = &\int W(\varepsilon) dV_0 \\
&+ \int \left\{ \mathbf{f}^a (\mathbf{e}_a - \nabla_a \mathbf{R}) + \mathbf{f}_0^a (\mathbf{e}_{0a} - \nabla_a \mathbf{R}_0) \right\} dV \\
&+ \int \left\{ \lambda^{ab} (g_{ab} - \mathbf{e}_a \cdot \mathbf{e}_b) + \lambda_0^{ab} (g_{0ab} - \mathbf{e}_{0a} \cdot \mathbf{e}_{0b}) \right\} dV \\
&+ \int \left\{ P \left( \frac{\sqrt{g}}{\sqrt{g_0}} - \beta(\varepsilon) \right) \right\} dV + \int \left\{ \lambda_\varepsilon \left( \frac{\sqrt{\hat{g}}}{\sqrt{\hat{g}_0}} - (1+\varepsilon) \right) \right\} dV,
\end{aligned} \quad (A1)$$

where $a, b = 1, 2, 3$, $dV_0 = \sqrt{g_0} dx_1 dx_2 dx_3$, $dV = \sqrt{g} dx_1 dx_2 dx_3$. $H$ is a function of $\varepsilon$, $\mathbf{e}_a$, $\mathbf{e}_{0a}$, $\mathbf{R}$, $\mathbf{R}_0$, $g_{ab}$, $g_{0ab}$, the definitions of which are captured by the Lagrange multipliers $\mathbf{f}^a$, $\mathbf{f}_0^a$, $\lambda^{ab}$, $\lambda_0^{ab}$, $P$, $\lambda_\varepsilon$. For brevity, we omit the explicit dependence of $\beta$ and $W$ on $x_3$. The Lagrange multiplier $\mathbf{f}^a$ corresponds to the stress tensor $\Sigma$ via the relation $\Sigma = -\mathbf{f}^a \otimes \mathbf{e}_a$, where $\otimes$ is an outer product [77,78]. To find $\mathbf{f}^a$, we perform the variation with respect to $\varepsilon$, $\mathbf{e}_a$, $\mathbf{e}_{0a}$, $g_{ab}$, $g_{0ab}$ and obtain the following system of equations:

$$\varepsilon : -P\beta(\varepsilon)\frac{\partial \beta(\varepsilon)}{\partial \varepsilon} - \beta(\varepsilon)\lambda_\varepsilon + \frac{\partial W(\varepsilon)}{\partial \varepsilon} = 0,$$

$$\mathbf{e}_{0a} : -2\lambda_0^{ab}\mathbf{e}_{0b} + \mathbf{f}_0^a = 0, a,b = 1,2,3,$$

$$\mathbf{e}_a : -2\lambda^{ab}\mathbf{e}_b + \mathbf{f}^a = 0, a,b = 1,2,3,$$

$$g_{03a} : \left(-\beta(\varepsilon)^2 P + W(\varepsilon)\right) g^{03a} + 2\lambda_0^{3a}\beta(\varepsilon) = 0, a = 1,2,3,$$

$$g_{0ab} : -\beta(\varepsilon)\lambda_\varepsilon(1+\varepsilon)g^{0ab} + \left(-\beta(\varepsilon)^2 P + W(\varepsilon)\right)g^{0ab}$$

$$+2\lambda_0^{ab}\beta(\varepsilon) = 0, a,b = 1,2,$$

$$g_{3a} : P\beta(\varepsilon)g^{3a} + 2\lambda^{3a} = 0, a = 1,2,3,$$

$$g_{ab} : \lambda_\varepsilon(1+\varepsilon)g^{ab} + P\beta(\varepsilon)g^{ab} + 2\lambda^{ab} = 0, a,b = 1,2,$$

(A2)

where each equation corresponds to the variation with respect to the variable written on the left. For the variation with respect to the components of the metric tensors, we used the relations: $\partial g / \partial g_{ab} = gg^{ab}$, $\partial g_0 / \partial g_{0ab} = g_0 g_0^{ab}$, $a,b = 1,2,3$; $\partial \hat{g} / \partial g_{ab} = \hat{g}g^{ab}$, $\partial \hat{g}_0 / \partial g_{0ab} = \hat{g}_0 g_0^{ab}$, $a,b = 1,2$. We omitted $\lambda^{a3}$ due to the symmetry of $\lambda^{ab}$, $a,b = 1,2,3$. Note that in general $\partial \hat{g} / \partial g_{ab} \neq \hat{g}g^{ab}$, but in our case $\partial \mathbf{R} / \partial x_3$ is always $\perp \partial \mathbf{R} / \partial x_a$, $a = 1,2$, as we do not consider tilt deformations. From Eqs. (10), we see that the variations with respect to the variables with index "0" are decoupled from the rest ones.

Now, we disentangle the system of equations (10) step by step in the following way:

$$\lambda_\varepsilon = \frac{\frac{\partial W(\varepsilon)}{\partial \varepsilon} - P\beta(\varepsilon)\frac{\partial \beta(\varepsilon)}{\partial \varepsilon}}{\beta(\varepsilon)},$$

$$\lambda^{ab} = \frac{g^{ab}\left(P\beta(\varepsilon)(1+\varepsilon)\frac{\partial \beta(\varepsilon)}{\partial \varepsilon} - (1+\varepsilon)\frac{\partial \beta(\varepsilon)}{\partial \varepsilon} - \beta(\varepsilon)^2 P\right)}{\beta(\varepsilon)},$$

$$\lambda^{3a} = -\frac{P\beta(\varepsilon)g^{3a}}{2}, a = 1,2,3,$$

(A3)

$$\mathbf{f}^a = \frac{-\frac{\partial W(\varepsilon)}{\partial \varepsilon}(1+\varepsilon) - P\beta(\varepsilon)^2 + P\beta(\varepsilon)(1+\varepsilon)\frac{\partial \beta(\varepsilon)}{\partial \varepsilon}}{\beta(\varepsilon)}\mathbf{e}^a,$$

$$\mathbf{f}^3 = -P\beta(\varepsilon)\mathbf{e}^3,$$

where $\mathbf{e}^a \equiv g^{ab}\mathbf{e}_b, a,b = 1,2,3$. Recall that for brevity we have so far omitted the explicit dependence of $W$ and $\beta$ on $x_3$. Now, choosing the $x_3$-axis to coincide with the previously introduced $z$-axis, we can construct the stress tensor:

$$\Sigma(\varepsilon,z) \equiv -\mathbf{f}^a \otimes \mathbf{e}_a$$

$$= \frac{\left\{\frac{\partial W(\varepsilon,z)}{\partial \varepsilon}[1+\varepsilon] + P\beta(\varepsilon,z)^2 - P\beta(\varepsilon,z)(1+\varepsilon)\frac{\partial \beta}{\partial \varepsilon}(\varepsilon,z)\right\}}{\beta(\varepsilon)}(\mathbf{1} - \mathbf{N} \otimes \mathbf{N})$$

(A4)

$$+P\beta(\varepsilon,z)\mathbf{N} \otimes \mathbf{N},$$

where $\mathbf{1}$ is the unit matrix and $\mathbf{N}$ the unit normal along the $z$-axis. $\Sigma$ acts in the following way: an outward unit normal to some region gives the force per unit area onto this region. In equilibrium the normal force is constant, i.e., $P\beta(\varepsilon,z) = \text{const} = -P_z$, where $P_z$ is the value of the normal pressure. Inserting $P\beta(\varepsilon,z) = -P_z$ into Eq. (10), we obtain:

$$\Sigma(\varepsilon,z) \equiv -\mathbf{f}^a \otimes \mathbf{e}_a$$
$$= \frac{\left\{\dfrac{\partial W(\varepsilon,z)}{\partial \varepsilon}[1+\varepsilon] - P_z\beta(\varepsilon,z) + P_z(1+\varepsilon)\dfrac{\partial \beta}{\partial \varepsilon}(\varepsilon,z)\right\}}{\beta(\varepsilon,z)}(\mathbf{1} - \mathbf{N} \otimes \mathbf{N}) \quad (A5)$$
$$-P_z \mathbf{N} \otimes \mathbf{N}.$$

The force factor is by definition equal to $\dfrac{\partial W(\varepsilon,z)}{\partial \varepsilon}$ and we now can express it from $\Sigma$ as:

$$F(\varepsilon,z) \equiv \frac{\partial W(\varepsilon,z)}{\partial \varepsilon} = \frac{(S(\varepsilon,z,P_z) + P_z)\beta(\varepsilon,z)}{1+\varepsilon} - P_z \frac{\partial \beta(\varepsilon,z)}{\partial \varepsilon}, \quad (A6)$$

where $S(\varepsilon,z,P_z)$ is the lateral stress, i.e., the lateral part of $\Sigma$.

**B. Second derivative of $\beta(\varepsilon,z)$**

In this appendix, we provide the expression and measured value for $\left.\dfrac{\partial^2 \beta(\varepsilon,z)}{\partial \varepsilon^2}\right|_{\varepsilon=0}$. Taking the derivative of Eq. (5) with respect to $P_z$ and then with respect to $\varepsilon$, we find:

$$\left.\frac{\partial^2 \beta(\varepsilon,z)}{\partial \varepsilon^2}\right|_{\varepsilon=0} = \frac{\partial}{\partial \varepsilon}\left\{\left(\frac{\partial S(\varepsilon,z,P_z)}{\partial P_z} + 1\right)\frac{\beta(\varepsilon,z)}{1+\varepsilon}\right\}\bigg|_{\varepsilon=0}. \quad (B1)$$

To find $\dfrac{\partial S(\varepsilon,z,P_z)}{\partial P_z}$, at a given fixed tension, corresponding to stretching $\varepsilon$, we varied $P_z$, setting its values to −50, −25, 1, 25, and 50 bar. The scaling function and $\beta(\varepsilon,z)$ at the right-hand side of Eq. (B1) were approximated by the linear relation $\beta(\varepsilon,z) = 1 + \gamma(z)\varepsilon$. The resulting values of $\left.\dfrac{\partial^2 \beta(\varepsilon,z)}{\partial \varepsilon^2}\right|_{\varepsilon=0}$ are shown in Figure B1.

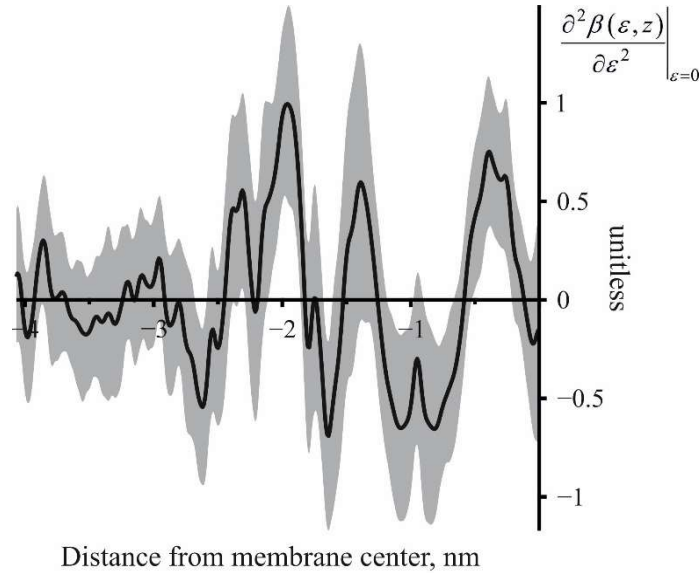

Distance from membrane center, nm

Figure B1. $\left.\dfrac{\partial^2 \beta(\varepsilon,z)}{\partial \varepsilon^2}\right|_{\varepsilon=0}$ as a function of the distance from the membrane center. The black curve and gray color are the mean value and 95% confidence band, respectively.

From Figure B1, it follows that the maximum absolute value of $\left.\dfrac{\partial^2 \beta(\varepsilon,z)}{\partial \varepsilon^2}\right|_{\varepsilon=0}$ do not exceed $\approx 1.5$. The moments $P_z \int_{m_0} \left.\dfrac{\partial^2 \beta(\varepsilon,z)}{\partial \varepsilon^2}\right|_{\varepsilon=0} (z-z_0)^n\, dz$ relative to the location of the neutral surface, $z_0 \approx -1.2$, at $P_z = 1$ bar, are $0.002 \pm 0.002$ $k_B T/\text{nm}^2$, $-0.0006 \pm 0.0037$ $k_B T/\text{nm}$ and $0.002 \pm 0.007$ $k_B T$ for $n = 0$, 1 and 2, respectively, and thus are negligibly small in comparison with the elastic parameters of DPPC.

## C. Measured values of elastic parameters

In this appendix, we provide the table of the measured elastic parameters.

TABLE C1. Values of the obtained monolayer elastic parameters: stretching modulus $k_A$, location of the neutral surface $z_0$, spontaneous curvature $K_{0,m}$, bending modulus $k_m$, and the deviation of the local Poisson's ratio from incompressibility condition $|\nu(z)-0.5|_{\text{Max}}$. $k_A$, $z_0$, $K_{0,m}$ and $k_m$ are measured using Eqs. (4a–c) under the assumptions of the local incompressibility (uniform scaling) and global incompressibility (exact scaling). The sign convention for spontaneous curvature: a positive value implies that the area of the head region tends to be larger than that of the tails. The errors indicate the standard deviation.

| | DPPC, T = 325 K | | DPPC/DOPC, 50/50, T = 325 K | | DOPC, T = 325 K | | DOPC, T = 300 K | |
|---|---|---|---|---|---|---|---|---|
| | Uniform scaling | Exact scaling | Uniform scaling | Exact scaling | Uniform scaling | Exact scaling | Uniform scaling | Exact scaling |
| $k_A$, $k_B T/\text{nm}^2$ | 32.1 ± 0.1 | 33.0 ± 0.2 | 33.9 ± 0.2 | 35.1 ± 0.2 | 35.6 ± 0.1 | 36.5 ± 0.2 | 43.4 ± 0.2 | 44.6 ± 0.2 |
| $z_0$, nm | −1.16 ± 0.01 | −1.19 ± 0.01 | −1.08 ± 0.01 | −1.11 ± 0.01 | −1.02 ± 0.01 | −1.04 ± 0.01 | −1.04 ± 0.01 | −1.06 ± 0.01 |
| $K_{0,m}$, nm$^{-1}$ | −0.18 ± 0.01 | −0.17 ± 0.01 | −0.31 ± 0.02 | −0.29 ± 0.01 | −0.32 ± 0.02 | −0.31 ± 0.02 | −0.32 ± 0.02 | −0.30 ± 0.02 |
| $k_m$, $k_B T$ (pN·nm) | 14.8 ± 0.6 (66.4 ± 2.7) | 15.3 ± 0.6 (68.9 ± 2.7) | 8.2 ± 0.4 (37.0 ± 2.0) | 9.0 ± 0.4 (40.2 ± 1.9) | 8.3 ± 0.6 (37.1 ± 2.5) | 8.7 ± 0.6 (39.2 ± 2.5) | 10.1 ± 0.8 (42.0 ± 3.2) | 10.8 ± 0.8 (44.8 ± 3.2) |
| $|\nu(z)-0.5|_{\text{Max}}$, unitless | — | 0.12 ± 0.01 | — | 0.09 ± 0.01 | — | 0.07 ± 0.01 | — | 0.08 ± 0.01 |

**ACKNOWLEDGMENTS**
The work was supported by the Ministry of Science and Higher Education of the Russian Federation and Russian Science Foundation grant #22-24-00661.